\documentclass{Interspeech2024}

\interspeechcameraready 

\makeatletter
\def\blfootnote{\gdef\@thefnmark{}\@footnotetext}
\makeatother

\usepackage{hyperref}
\usepackage{tablefootnote}
\hypersetup{
    colorlinks=true,
    linkcolor=blue,
    filecolor=magenta,      
    urlcolor=cyan,
    pdftitle={Overleaf Example},
    pdfpagemode=FullScreen,
    }
\usepackage{subcaption}
\title{Small-E: \\  Small Language Model with Linear Attention for Efficient Speech Synthesis}
\name{Théodor Lemerle, Nicolas Obin, Axel Roebel}
\address{
STMS Lab\\
  IRCAM, CNRS, Sorbonne Université \\
  Paris, France}
\email{}

\begin{document}

\maketitle

\begin{abstract}
Recent advancements in text-to-speech (TTS) powered by language models have showcased remarkable capabilities in achieving naturalness and zero-shot voice cloning. Notably, the decoder-only transformer is the prominent architecture in this domain. However, transformers face challenges stemming from their quadratic complexity in sequence length, impeding training on lengthy sequences and resource-constrained hardware. Moreover they lack specific inductive bias with regards to the monotonic nature of TTS alignments. In response, we propose to replace transformers with emerging recurrent architectures and introduce specialized cross-attention mechanisms for reducing repeating and skipping issues. Consequently our architecture can be efficiently trained on long samples and achieve state-of-the-art zero-shot voice cloning against baselines of comparable size. Our implementation and demos are available at \url{https://github.com/theodorblackbird/lina-speech}.
\end{abstract}

\vspace{-0.25cm}

\section{Introduction}

% NO : pour info l'introducgion ne doit pas dépasser la moitité de la première colonne de la deuxième page (pour une article de 4+1 pages)

% NO : Avant related works il faut un peu de contexte. Par exemple un rapide rappel de la synthèse TTS, de son évolution +/- récente, et des problèmes liés. Tu pourrais ici faire donc une description de l'histoire du TTS neuronal, de son évolution, et des tendances historiques (Eventuellement aussi de ses applications, mais je trouve ça pas super pertinent). 

\subsection{Context and Related Works}

\blfootnote{This research was supported by the project EXOVOICES ANR-21-CE23-0040 and funded by the French National Research Agency.}
Over the recent years, neural text-to-speech synthesis (TTS) has gained spectacular improvements in terms of quality with a diversity of approaches and paradigms \cite{ren2020fastspeech, kim2021conditional, le2023voicebox, guo2023voiceflow}. In particular, discrete speech and audio representations allowed immediate use of well-established decoder-only transformers such as GPT \cite{gpt2} in many state-of-the-art text-to-audio and text-to-speech model. However, transformers rely on the self-attention ‘‘time-mixing"\cite{rwkv4} operation which can be efficiently trained in parallel but suffers from quadratic complexity with respect to the sequence length.
The challenge of designing sequence modeling architecture that can compete with transformers has sparked a resurgence in research on recurrent neural networks (RNNs). This work introduces the broad term ‘‘linear attention" to denote this emerging class of RNNs that replaces self-attention for linear complexity ‘‘time-mixing" while keeping performances and high training throughput. %Specifically they show an important improvement over LSTM or GRU that do not show competitive results nor high training throughput against transformer at sequence modeling.  

%\subsection{Related works}

This paper primarily relates to speech models formulated as language models (LMs) or employing discrete audio codecs through Residual Vector Quantization (RVQ). VALL-E \cite{valle} employs an autoregressive transformer to predict the first quantizer and a parallel transformer for the residuals. Before the rise of RVQ codecs, Tortoise \cite{tortoise} achieved a significant improvement through scaling up and leveraged a decoder-only transformer to predict a VQ representation of the mel spectrogram. Some other works introduce semantic codes as low frame rate audio latents, following advancements in self-supervised speech representations.
For instance Bark \cite{bark} separately predicts semantic codes from text, first quantizers from semantic codes, and residuals with three decoder-only transformers. SoundStorm \cite{soundstorm} predicts audio from semantic codes in parallel by leveraging a MaskGit \cite{maskgit} architecture. In contrast, NaturalSpeech2 \cite{naturalspeech} avoids the language model formulation by learning the continuous latents of an RVQ codec with a diffusion model, sidestepping autoregressive modeling or semantic encoding and instead relying on given durations and fundamental frequency. 

%While forming a relatively new family of speech synthesis model, more traditional architecture in the lineage of FastSpeech2\cite{ren2020fastspeech} or VITS\cite{kim2021conditional}, that relies on duration and/or pitch prediction and mel spectrogram such as VoiceBox\cite{le2023voicebox} or VoiceFlow \cite{guo2023voiceflow} reaches state-of-the-art quality on various model sizes.

\vspace{-0.15cm}

\subsection{Linear surrogate of decoder transformer}

Unlike previous RNNs such as LSTM or GRU, transformers are significantly faster to train, do not suffer from vanishing gradient and demonstrate scalability with parameters reaching into the hundreds of billions. Further hardware-aware implementation \cite{dao2022flashattention} of self-attention has established it as a prevalent choice for sequence modeling, including applications in audio processing.
General softmax-based attention involves three sequences, denoted as \(\mathbf{Q} \in \mathbb{R}^{N \times d}\), \(\mathbf{K} \in \mathbb{R}^{N' \times d}\), and \(\mathbf{V} \in \mathbb{R}^{N' \times d'}\), along with an optional mask \( \mathbf{M} \in \mathbb{R}^{N \times N'}\). The attention function is defined as:
\vspace{-0.25cm}
\begin{equation}\label{eq:att}
\begin{aligned}
\text{Att}(\mathbf{Q}, \mathbf{K}, \mathbf{V}) = \text{softmax}(\frac{\mathbf{Q}\mathbf{K}^T}{\sqrt{d}} \odot \mathbf{M})\mathbf{V}.
\end{aligned}
\vspace{-0.25cm}
\end{equation}

When \(\mathbf{Q}\), \(\mathbf{\mathbf{K}}\), and \(\mathbf{\mathbf{V}}\) represent different linear projections of the same input sequence \(\mathbf{X}\ \in \mathbf{R}^{N\times d} \) (and are therefore function of $\mathbf{X}$), the resulting function \( \mathbf{X} \mapsto \text{Att}(\mathbf{Q}, \mathbf{K}, \mathbf{V})\) is referred to as self-attention. Backing the success of GPT2 \cite{gpt2} and successors for natural language modeling, the decoder-only transformer architecture can be generalized with the terminology proposed in \cite{rwkv4}:
\vspace{-0.15cm}
\begin{equation}
\begin{aligned}\label{eq:gpt}
\mathbf{Y'} = \mathbf{X} + \text{TimeMixing}(\text{Norm}(\mathbf{X})), \\
\mathbf{Y} = \mathbf{Y'} + \text{ChannelMixing}(\text{Norm}(\mathbf{Y'})).
\end{aligned}
\vspace{-0.15cm}
\end{equation}
Initially built with self-attention for $\text{TimeMixing}$ \cite{attention}, position-wise feed-forward network for $\text{ChannelMixing}$ and layer normalization for $\text{Norm}$.
In the context of autoregressive modeling, a causal mask \(\textbf{M}_{i,j} = \textbf{1}_{i<j}\) is employed to prevent tokens from attending to relative future tokens, enabling the model to function as an autoregressive model while being trained in parallel. 

\begin{figure}
\vspace{-0.5cm}
\centering
\begin{subfigure}{0.4\textwidth}
    \includegraphics[width=\textwidth, trim=0 0 0 1cm]{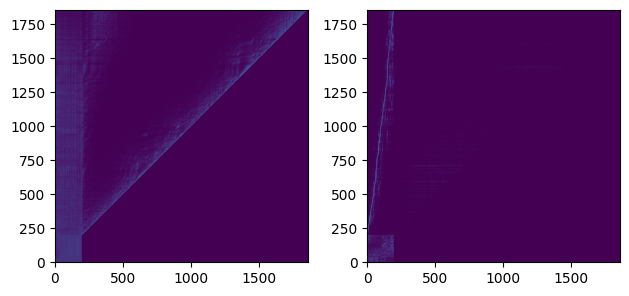}
    \caption{In decoder-only LM TTS models \cite{valle, bark, metavoice}, attention scores either boil down to cross-attention or local reasoning so that a large proportion of tokens are not attended.}
    \label{fig:second}
\end{subfigure}
\begin{subfigure}{0.4\textwidth}
    \includegraphics[width=\textwidth]{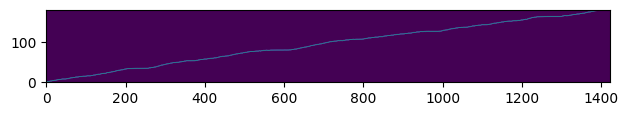}
    \caption{In our work, we only use two layers of cross-attention compared to self-attention in every layer.}
    \label{fig:first}
\end{subfigure}
%\hfill
%\hfill
\caption{Decoder-only attention weight tend to behave as an encoder-decoder.}
\vspace{-0.55cm}
\label{fig:figures}
\end{figure}

However, self-attention exhibits quadratic complexity concerning sequence length during training and inference. Simultaneously, it has been observed that in certain scenarios, self-attention tends to focus on local reasoning \cite{parcollet2023sumformer}, as evidenced in text-to-speech by almost diagonal attention weights in practice (see \autoref{fig:figures}). This observation suggests that computing every pair-wise relation in \(\mathbf{Q}\mathbf{K}^T\) at every layer may not be essential. 
These insights align with recent developments in recurrent architectures which have emerged as potential replacements for transformers in natural language modeling. For instance, RWKV \cite{rwkv4} is an RNN designed as an alternative to transformers. RWKV incorporates new mechanisms for both TimeMixing (known as ‘‘WKV" and is closely linked to some form of linear attention \cite{kitaev2020reformer, shen2019efficient, zhai2021attention}) and ChannelMixing (involving linear interpolation of current and past token). In the lineage of State-Space Model \cite{s4}, Mamba \cite{mambassm} unifies TimeMixing and ChannelMixing operations in (\autoref{eq:gpt}), removes the linear time-invariant assumption with data-dependency and introduces parallelization through parallel scan. RetNet \cite{retnet} features linear attention with decaying state for TimeMixing allowing efficient chunkwise computation. Gated Linear Attention (GLA) \cite{gated} explores hardware efficient chunkwise form of linear attention with data-dependent transition.
All offer alternatives to the original transformer decoder-only block for language modeling with competitive throughput and performance for language modeling as demonstrated on various tasks.
They scale linearly with the sequence length during training, opening the door to training on long sequences thus capturing long-term dependencies at a lower cost. We refer to them as \textbf{Linear Causal Language Model (LCLM)} blocks. To the best of our knowledge their usage for audio generative modeling remains largely unexplored.

\subsection{TTS as conditional codec language modeling}
The remarkable ability of language models to adapt from unseen inputs sample is known as ‘‘in-context learning" \cite{lee2023exploring}, and has been successfully adapted for TTS for zero-shot voice continuation \cite{valle}. For text-to-speech we follow conditional codec language modeling formulation as introduced by \cite{valle}, given $\mathbf{x} = \{x_0, \ldots, x_N\}$ a text transcription, $\mathbf{y} \in \{1, \ldots, C\}^{Q \times T}$ a RVQ representation of the corresponding audio with $Q$ quantizers of codebook size $N_c$  we formulate it as 
\begin{equation}
p(\mathbf{y}|\mathbf{x}) = \prod_{t=0}^T \prod_{q=1}^{Q} p(\mathbf{y_{q,t}}|\mathbf{x}, \mathbf{y_{<q,  \leq t}}).  
\end{equation}
By concatenating source and target transcriptions and by providing only source audio tokens during inference, conditional codec modeling turns into a zero-shot voice cloning model without explicit need of speaker encoder module. In contrast with natural language modeling and because of the hierarchical nature of RVQ, we must account of the conditional dependencies between succeeding residuals. Previous work \cite{valle, bark} tend to train separate models for first ‘‘coarse" quantizers to ‘‘finer" successive residuals.

\subsection{Positioning and contributions}

\vspace{-0.15cm}

\textcolor{black}{
Being able to train on large datasets is a crucial aspect for diverse and expressive speech generation. Typical LM are difficult to train in the limited hardware regime (\textit{ie} consumer grade GPU). Our model shows that language modeling can be successfully adapted to the small model regime with careful architecture considerations :
\begin{itemize}
    \item This paper introduces {\bf Linear Causal Language Model (LCLM)} blocks instead of autoregressive transformers commonly used in language modeling for audio application. To the best of our knowledge this is the first time they are used for text-to-speech. As a consequence we are able to train efficiently on long samples (up to 30s). We hypothesize that it is crucial for learning expressive speech.
    \item We introduce a {\bf Position-Aware Cross-Attention (PACA)} mechanism which is specifically designed for text-to-speech and helps with skipping and repeating issues.
    \item The proposed model has competitive performance on zero-shot voice cloning TTS by comparison to existing TTS models of the same size, while requiring much less resources during training.
\end{itemize}
}

\vspace{-0.35cm}

\section{Small-E}

\begin{figure}
\vspace{-0.55cm}
\centering
\includegraphics[scale=0.8, trim=1cm 0.70cm 1cm 1cm]{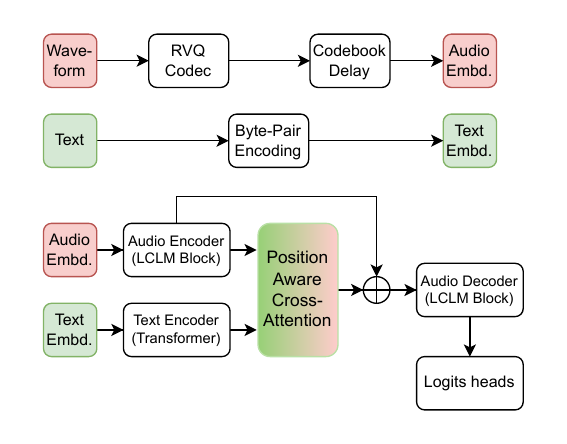}
\caption{Model Overview. Top: input pipeline. Bottom: encoder-decoder architecture. $\bigoplus$ means summation.}
\vspace{-0.5cm}
\label{fig:model_overview}
\end{figure}

This section presents Small-E, a multi-speaker neural TTS with zero-shot voice cloning capabilities. Small-E belong to the family of neural codec language model such as \cite{valle, bark, metavoice, soundstorm}. 
In contrast with previous TTS codec LM model that leverages decoder-only (GPT) transformers, Small-E relies on encoder-decoder architecture. Indeed, we observed that previous decoder-only transformers tend to behave internally as an encoder-decoder (see \autoref{fig:figures}) leading to a potential waste of compute. 
\textcolor{black}{The general architecture is presented in Figure \ref{fig:model_overview}, text is encoded through a non-causal transformer, audio is encoded with a stack of LCLM blocks. Both encoders outputs are fed to the cross-attention that learns to align text to audio. The audio decoder (same as the audio encoder) takes audio embeddings and cross-attention output. The decoder output is projected to logits.}
%in order to train efficiently on long utterances and observe that two cross-attention layers is sufficient for learning a mapping between text and audio.
%
%
%

\subsection{Model architecture}

The input pipeline for audio and text compression is processed in the following manner.
Audio is compressed with an RVQ codec. We employ a codebook delaying scheme introduced by MusicGen \cite{copet2023simple} in order to enforce the conditional dependencies between residual codebooks.
Text is compressed with byte-pair encoding before embedding. Then the text embedding is processed with an non-causal transformer encoder, the audio embeddings are processed with an encoder consisting of a stack of LCLM blocks. Both encoder outputs are then fed to a Position-Aware Cross-Attention, the output being the text embedding attended for each audio embedding. The text embedding and audio embedding are then superposed and fed to an audio decoder similar to the audio encoder.
The three basic blocks are: \\ %\vspace{0.1cm}
\noindent \textbf{Text Encoder} This component comprises a stack of non-causal (parallel) transformer encoders with RoPE positional embedding \cite{su2023roformer}.

\noindent \textbf{Audio Encoder/Decoder} We investigated various LCLM blocks including RWKV \cite{rwkv4} v5.2/v6, Mamba \cite{mambassm} and GLA \cite{gated}. In early experiments we found that they offer comparable performances.

\noindent \textbf{Position-Aware Cross-Attention (PACA)} Autoregressive speech modeling is prone to skipping and repeating issues \cite{valle, tacotron2, flowtron}. We introduce a simple tweak to enforce position awareness.
In the conventional formulation of cross-attention between text and audio, represented as:
\vspace{-0.2cm}
\begin{equation}
\mathbf{Y} = \text{Att}(\mathbf{Q}, \mathbf{K}, \mathbf{V}),
\vspace{-0.2cm}
\end{equation}
where $\mathbf{Q}$ is the audio latent sequence, and $\mathbf{K}$, $\mathbf{V}$ are linear projections of the text latent sequence, attention is computed independently for every time step, without considering previous attended text latent.

\begin{figure}[h!]
\centering
\includegraphics[scale=0.6, trim={0 0 0 2cm}]{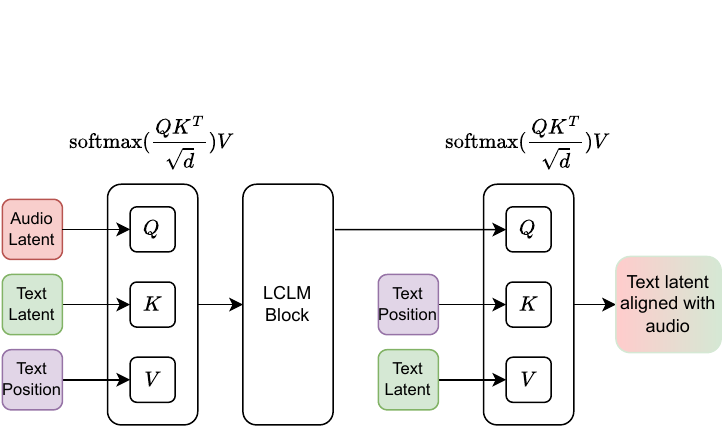}
\caption{Position-Aware Cross-Attention}
\label{fig:position_aware_cross_attention}
\vspace{-0.25cm}
\end{figure}

To address this limitation, we propose a modification to the cross-attention mechanism by explicitly materializing position information along with a feedback loop to propagate past positions (see \autoref{fig:position_aware_cross_attention}). Firstly, a cross-attention is computed by selecting text positions only:
\vspace{-0.1cm}
\begin{equation}
\mathbf{Y^{(1)}} = \text{Att}(\mathbf{Q}, \mathbf{K}, \mathbf{P}),
\vspace{-0.1cm}
\end{equation}
where $\mathbf{P}$ is sinusoidal positional embedding along the text latent positions, that is:
\vspace{-0.5cm}
\begin{equation}
\begin{aligned}
\mathbf{P_{t, 2d}} = \sin(t/10000^{2d/d_b}), \\
\mathbf{P_{t, 2d+1}} = \cos(t/10000^{2d/d_b}).
\end{aligned}
\vspace{-0.1cm}
\end{equation}

This ensures that $\mathbf{Y^{(1)}}$ contains only information about the position of the attended text latent, rather than actual text content. These positions are then fed into a LCLM block represented by the transition function $f$ to introduce a feedback loop on the positions:
\begin{equation}
\vspace{-0.25cm}
\mathbf{Y^{(2)}_{t+1}, H_{t+1} = f(Y^{(1)}_{t}, H_t)}.
\vspace{0.2cm}
\end{equation}
This causal linear LCLM block can be any recurrent LM block such as RWKV \cite{rwkv4}, Mamba \cite{mambassm} or GLA \cite{gated}. \vspace{0.1cm}

Finally, a cross-attention of $\mathbf{Y_2}$ against $\mathbf{P}$ is performed to select $\mathbf{V}$ (containing text information), mapping the position $\mathbf{Y^{(2)}}$ to the text latent:
\vspace{-0.15cm}
\begin{equation}
\mathbf{Y^{(3)}} = \text{Att}(\mathbf{Y^{(2)}}, \mathbf{P}, \mathbf{V}).
\vspace{-0.1cm}
\end{equation}
It is important to note that the positional embedding $\mathbf{P}$ is not superposed onto any latent vector but rather materialized independently, constraining the model to accurately encode positional information to effectively attend to the text content. We set $d_b$ to be significantly smaller than the model dimension ($d_b \leq 64$) to keep additional operations negligible.
This approach is reminiscent of Location Sensitive Attention \cite{tacotron2}, while being an order of magnitude faster due to the efficiency of LCLM blocks.
The optimization objective is cross-entropy loss between original RVQ codec and logits prediction.

\vspace{-0.25cm}

\section{Experimental Evaluation}

\subsection{Dataset}

Small-E was trained on Librilight medium \cite{librilight}, consisting of approximately 5,000 hours of multi-speaker English speech recordings reading audio books, collected from LibriVox. We used the provided recipes to get samples of approximately 25 seconds. Speech utterances were transcribed textually using Ocotillo \cite{ocotillo} speech recognition system. Validation set is made of 2,000 random utterances. For evaluation, the proposed model and the benchmark models (see below) were compared on the LibriTTS test split \cite{librilight}, in particular by insuring training and testing speakers do not overlap.

\vspace{-0.15cm}

\subsection{Implementation details}
\label{implementation-details}

For the text encoding, the proposed model used byte-pair encoding with a vocabulary size of 256 computed on the dataset transcription.
For the audio encoding we used EnCodec \cite{encodec} at 3kbps bitrate. 
In a preliminary experiment, the proposed model has been compared with different LCLM blocks, including RWKV, Mamba, and GLA.  
For each, the text encoder consists of 9 layers of non-causal transformer, each layer of dimension 512 with 8 heads. For the experiments, we chose Gated Linear Attention as LCLM Block, each block consisting of 6 layers of inner dimension 512 with 2 heads. 
We observed during this preliminary that GLA is performing similarly as Mamba and RWKV in terms of validation loss while giving slightly better training throughput with respect to our configuration (i.e fixed batch size and number of parameters). 
For this reason, the proposed model is using GLA blocks in the remaining of this paper. %Nevertheless, the provided implementation is sufficiently generic to handle the compared LCLM blocks and trained model are available for each of them (RWKV, Mamba, and GLA).
The whole model consists of 64M trainable parameters. 
For the training, Adam optimizer was used with a learning rate equal to $5 
\mathrm{e}{-4}$, with momentum $\beta_1 = 0.9$, $\beta_2 = 0.999$, and weight decay of $0.1$. We group sentences of similar length within 10 buckets and use dynamic batch size with target size of approximately $80,000$ audio tokens. We used gradient clipping of $1.0$. Trainings were done on 4 RTX3080 (10GB VRAM each) during two days, consisting of 15 epochs over the dataset. During inference we use top-$k$ sampling with $k$ set to $100$ for the first quantizer and greedy decoding for the residuals.

\vspace{-0.15cm}

\subsection{Benchmark}

We compared Small-E with YourTTS \cite{yourtts} which is a common baseline for the evaluation of multi-speaker TTS models (e.g., \cite{valle, naturalspeech}). For the comparison, we used the official checkpoint which has been trained on a multilingual dataset comprising VCTK (English), LibriTTS (English), and Portuguese split of MLS.
In addition, we also compared with MetaVoice \cite{metavoice} as a strong baseline, an open-source and open-weight GPT model of 1.2B parameters trained on 100k hours of speech from a private dataset. This constitutes to our knowledge the strongest codec language model TTS model publicly available.  Notably for decoding EnCodec tokens, our model rely on Vocos \cite{vocos} at 3kbps. This is in contrast with MetaVoice which leverages MultiBand diffusion at 6kbs and an additional post-net. We regret that most of the baselines belonging to our family don't have publicly available official implementations, limiting our subjective and objective evaluation.

\vspace{-0.25cm}

\subsection{Methodology}

\vspace{-0.1cm}

\subsubsection{Objective evaluation}

As for the quantitative objective evaluation, we investigated the performance of the proposed Small-E architecture in terms of training throughput, i.e., the throughput measured by means of audio-tokens per seconds and the perplexity of the LM as an indicator of the reconstruction error as the exponential of the cross-entropy loss. This is measured on Librilight medium with the setup described in Section \ref{implementation-details}. 
Additionally, we conducted an ablation study to investigate the role of the proposed PACA mechanism with respect to the skips and repetitions problem known as a common issue of auto-regressive models \cite{valle, tacotron2, regotron}. To do so, we followed the methodology presented in \cite{regotron}.
100 utterances were randomly picked up from the validation set and were manually inspected in terms of skips and repetitions by comparison of the reference utterance.
This methodology is preferred to the common measurement of the word error rate since the skip and repetition problem is specific to auto-regressive models \cite{regotron, valle}. 

\vspace{-0.25cm}

\subsubsection{Subjective Evaluation}

A subjective evaluation was additionally conducted to assess the {\em naturalness} and the {\em similarity} to the reference speaker of the considered speech sample. 
The experiment consisted in presenting to the participants a speech sample and a reference speech sample of the same speaker but pronouncing another utterance.
The participants were asked to judge the speech sample with respect of the following instructions on a 5-degree MOS scale : (1) {\em naturalness}: {\em to which extent the speech sample is judged as natural as real human speech?};
%(‘‘bad", ‘‘poor", ‘‘fair", ‘‘good", and ‘‘excellent"). 
%Naturalness is solely judged on the basis of the provided speech sample.
(2) {\em similarity }: to the reference speaker:  {\em to which extent the speech sample is judged close to the reference speaker?} 
%(‘‘very dissimilar", ‘‘fairly dissimilar", ‘‘slightly similar", ‘‘fairly similar", and ‘‘very similar"). 
%
For each participant subject, an experiment run consisted into the judgement of 15 samples. These samples were randomly selected among a total of 50 utterances (the same for all models) $\times$ 4 models (the three models being and the real speech = 200 speech samples. 
The real speech was presented as a positive anchor to the participant. 
The whole experiment has been conducted using the Prolific platform with a mix of 70 native and non-native English speakers.

\vspace{-0.3cm}

\section{Results and Discussion}

Table \ref{tab:obj-throughput} presents the training throughput of Small-E with comparison to a standard decoder-only architecture (following \cite{valle} implementation) taken as a baseline of the LMs generative family. Small-E training is significantly faster compared to this baseline architecture with same amount of parameters, with a relative increase of 62 \%. This comes with a slight improvement of the perplexity, indicating that training throughput gain doesn't come at the cost of performance.

\begin{table}[h!]
\scriptsize
  \caption{Training throughput. Throughput is measured by means of audio token per second (in kilo tokens per second), and perplexity (referred to as ppl).}
    \label{tab:obj-throughput}
  \centering
  \begin{tabular}{ c c c }
    \toprule
    \multicolumn{1}{c}{\textbf{Model}} & 
                                        
                                         \multicolumn{1}{c}{\textbf{Audio token per second (kT/s)} \tablefootnote{Using EnCodec\cite{encodec} at 3kbps with delaying scheme\cite{copet2023simple},  it consists of 4 tokens generated in parallel per decoding step.} $\uparrow$}
                                        & \multicolumn{1}{c}{\textbf{ppl} $\downarrow$} \\
    \midrule
    Small-E            &      316     & 18.33             \\
    Decoder-only (GPT)         &     195      & 19.68  \\

    \bottomrule
  \end{tabular}
  \vspace{-0.5cm}
\end{table}

\vspace{0.25cm}

Table \ref{tab:obj-ablation} presents the results of the ablation study.
On the 100 generated utterances which were manually inspected, the Small-E version with PACA presents a drastic reduction of this problem, either in terms of skips or repetitions. This proves the efficiency of the proposed cross-attention mechanism as a solution to the skip and repetition problem.

\vspace{-0.15cm}

\begin{table}[h!]
\scriptsize
  \caption{Position-Aware Cross-Attention impact on 100 utterances. Number of utterances that contains at least one skip/repetition.}
  \label{tab:obj-ablation}
  \vspace{-0.1cm}
  \centering
  \begin{tabular}{ c c c }
    \toprule
    \multicolumn{1}{c}{\textbf{Model}} & 
                                        
                                         \multicolumn{1}{c}{\textbf{Skip} $\downarrow$}
                                        & \multicolumn{1}{c}{\textbf{Repeat} $\downarrow$} \\
    \midrule
    Small-E w. PACA           &      1     & 1             \\
    \hspace{0.3cm} w/o PACA         &     5      & 9  \\

    \bottomrule
  \end{tabular}
    \vspace{-0.1cm}
\end{table}

Table \ref{tab:subj-exp} presents the MOS scores obtained for the subjective evaluation. Under $t$-test, we found every pair of candidates to be significantly different ($p < 0.05$).
Firstly, we observe that Vocos at low bitrate presents a slight but significant degradation with comparison to the original speech sample.
Secondly, Small-E presents significantly higher scores than the baseline YourTTS both in terms of naturalness and similarity. Finally, Small-E presents significantly lower score than the strong baseline MetaVoice as expected since it consists of $20$ times more data and parameters, excluding it from training on limited hardware.

\vspace{-0.1cm}

\begin{table}[th]
\scriptsize
  \caption{Subjective evaluation. MOS for naturalness, and SMOS for similarity to the reference speaker.}
  \label{tab:subj-exp}
  \centering
  \begin{tabular}{ c c c c c c }
    \toprule
    \multicolumn{1}{c}{\textbf{Model}} & 
                                        
                                         \multicolumn{1}{c}{\textbf{Params.}}
                                        & \multicolumn{1}{c}{\textbf{MOS} $\uparrow$}
                                        & \multicolumn{1}{c}{\textbf{SMOS} $\uparrow$}\\
    \midrule
    Original            &           & $4.55 \pm 0.20$  & $4.62 \pm 0.23$~~~            \\
    Vocos 3kbps         &           & $4.27 \pm 0.21$  & $4.43 \pm 0.22$~~~             \\ \midrule
    Small-E (ours)      & 64M       & $3.16 \pm 0.28$  & $3.08 \pm 0.30$~~~    \\
    YourTTS             & 86M       & $2.56 \pm 0.24$  & $2.54 \pm 0.24$~~~           \\
    MetaVoice           & 1.2B      & $3.80 \pm 0.28$   & $3.91 \pm 0.28$~~~            \\

    \bottomrule
  \end{tabular}
  
\end{table}

\vspace{-0.55cm}

\section{Conclusion}
In this paper we presented Small-E, a TTS model based on codec language model. 
The proposed model tackles limitations of current LM TTS models. Firstly, we introduced Linear Causal Language Model in place of the traditional decoder-only transformer. 
Secondly, we introduced a cross-attention mechanism designed specifically to handle text and speech modalities in the context of TTS, with the idea of preventing the skip and repetition problem of auto-regressive models.
In contrast with existing work, we were able to show that training LM TTS model is interesting even on limited hardware and leads to state-of-the-art quality against model of the same size.
Experimental evaluation demonstrated the efficiency of the proposed model either in terms of training throughput, skip and repetition reduction, as well as naturalness and similarity to the reference speaker of the generated speech.
These observations constitute encouraging results opening the way for small and efficient generative TTS models. In future work we are interested in streaming TTS, taking advantage of the linear complexity of LCLM for very long or embedded synthesis.

\clearpage
\bibliographystyle{IEEEtran}
\bibliography{mybib}

\end{document}